\documentclass[a4paper,10pt, twocolumn, preprint,5p]{elsarticle}
\usepackage{lineno,hyperref}

\usepackage{graphicx}
\usepackage{epstopdf}
\usepackage{graphics}
\usepackage{amsmath}
\usepackage{threeparttable}
\usepackage{booktabs}
\usepackage{multirow}
\usepackage{color}
\usepackage{amssymb}
\usepackage{balance}
\usepackage{subfig}
\usepackage{tabularx}

\journal{Microprocessors and Microsystems}

\begin{document}

\begin{frontmatter}

%\title{Elsevier \LaTeX\ template\tnoteref{mytitlenote}}

\title{On Fault-Tolerant Design of Exclusive-OR Gates in QCA}

%\address{ISI}
\author{Dharmendra Kumar}
\author{Debasis Mitra}
\address{Department of Information Technology}
\address{National Institute of Technology, Durgapur-713209, India}
\author{Bhargab B. Bhattacharya}
\address{Advanced Computing and Microelectronics Unit}
\address{Indian Statistical Institute, Kolkata-700108, India}


%\ead{debasis.mitra@gmail.com}


\begin{abstract}
Design paradigms of logic circuits with Quantum-dot Cellular Automata (QCA) have been extensively studied in the recent past. Unfortunately, due to the lack of mature fabrication support, QCA-based circuits often suffer from various types of manufacturing defects and variations, and therefore, are unreliable and error-prone.
QCA-based Exclusive-OR (XOR) gates are frequently used in the construction of several computing subsystems such as adders, linear feedback shift registers, parity generators and checkers.
However, none of the existing designs for QCA XOR gates have considered the issue of ensuring fault-tolerance.
Simulation results also show that these designs can hardly tolerate any fault.
We investigate the applicability of various existing fault-tolerant schemes such as triple modular redundancy (TMR), NAND multiplexing, and majority multiplexing in the context of practical realization of QCA XOR gate. Our investigations reveal that these techniques incur prohibitively large area and delay and hence, they are unsuitable for practical scenarios. We propose here realistic designs of QCA XOR gates (in terms of area and delay) with significantly high fault-tolerance against all types of cell misplacement defects such as cell omission, cell displacement, cell misalignment and extra/additional cell deposition. Furthermore, the absence of any crossing in the proposed designs facilitates low-cost fabrication of such systems.

\end{abstract}


\end{frontmatter}

%\linenumbers

\section{Introduction}
\label{sec:intro}
\label{sec:Intro}
Having matured over around six decades, CMOS technology is now expected to reach its physical limit in the near future \cite{end_of_MooresLaw}. The never-ending quest for smaller computing devices has driven the research on alternative nanotechnologies. Quantum-dot Cellular Automata (QCA) \cite{lent_tnano1993} has emerged as a possible option in recent years. In QCA, information flows through basic elements (referred to as cells) not by actual flow of current, as in conventional CMOS based designs, but by coulombic interactions between electrons present in neighboring cells resulting in very low power dissipation \cite{lent_tnano1993}. Other promising features of QCA technology include high device packing density, high speed (in order of THz), inherent pipelining \cite{lent_jap1994}. Implementations of basic logic devices in QCA have been demonstrated \cite{lent_jap1994}. Design and simulation of common digital modules (both combinational and sequential) have been studied extensively \cite{Cho_tc2009, Lim_ICCAS_2012, swartzlander_asilomar2012, Purkayastha_mm_2016, Angizi_mm2015, Venkataramani_CNANO_2008}.

As in other nanotechnologies, QCA-based circuits often suffer from various types of manufacturing defects \cite{Lombardi_dfvlsi2008,momenzadeh_pdps2004,tahoori_tnano2004}. Experimental studies revealed that {\it cell misplacement} defects are the most common among all such defects \cite{tahoori_tnano2004}. Several types of cell misplacement defects viz. {\it cell displacement}, {\it cell misalignment}, {\it cell omission} have been reported in the literature so far \cite{tahoori_tnano2004}. Hence, low-cost fault-tolerant designs for such defects are needed.
Fault-tolerance and thermal characteristics of fundamental QCA logic devices have been analysed by Anduwan {\it et al.} \cite{tougaw_jap2010}. Errors due to random clock shifts in QCA circuits have been studied by Karim {\it et al. \cite{Karim_JETTA_2009}}.
Ma {\it et al.} \cite{Lombardi_dfvlsi2008} presented a comparative study on the applicability of a few generic fault-tolerant schemes such as triple modular redundancy (TMR) \cite{LyonsTMR1962}, NAND multiplexing \cite{neumann_1956}, and  majority multiplexing \cite{roy_tnano2005} in the context of reliable realization of QCA systems.
A few new fault-tolerant QCA designs of majority gates and adders have been reported in the literature \cite{Farazkish_mm2015,lombardi_date2006,Mitra_MJ_2016,Lombardi_cs2005,Bibhash_indicon2014}.  

An Exclusive-OR (XOR) gate is a digital logic gate that results a true output (logic 1) if one, and only one, of the two inputs of the gate is true. 
The associative nature of the exclusive-OR function implies the possibility of using exclusive-OR gates with three or more inputs. The exclusive-OR operation with three or more variables can be defined as an {\it odd function} where the output assumes logic value 1 if an odd number of variables be equal to 1 \cite{Mano_DigitalBook}.
XOR gates are often considered as important modules in digital circuit design due to their frequent use in the construction of several computing subsystems such as adders, linear feedback shift registers, parity generators, parity checkers, decoders for error correction and channel codes \cite{Mano_DigitalBook}.
An XOR gate with three or more inputs is usually constructed by appropriately connecting two or more than 2-input XOR gates only. However, as described in \cite{Feinstein_RMW2007}, XOR gates with four or higher inputs can be efficiently constructed with the help of 2-input and 3-input XOR gates.

Several designs for 2-input QCA XOR gates have been presented in recent past \cite{bhat_INDIACom2015,beigh_cs2013,hema_waset2011,shah_IOSR-JCE2014,beigh_ijpap2013,pallavi_ijsrd2014}. A multilayer 2-input XOR design is also given as sample file in QCADesigner version 2.0.3 \cite{Walus_QCADesigner}. A few 3-input XOR gate designs have also been presented recently \cite{Kianpour_IJNN_2014,Shin_IJCA_2015}. But, both of them have used 2-input XORs and hence shows very poor performance in terms of common design parameters such as latency and area. Most importantly, to the best of our knowledge, none of the XOR designs (both 2-input and 3-input) considered fault-tolerance. Again, wire crossing (coplanar or multi-layer), inherent to some of these designs, make them difficult to realize in practice \cite{Chaudhary_ICCAD2005}.  

In this paper, we investigate the applicability of various existing fault-tolerant schemes such as TMR, NAND multiplexing, and majority multiplexing in the context of practical realization of QCA XOR gate. Our investigations reveal that the XOR-function realized under these fault-tolerant schemes requires a large number of cells thereby showing a very poor fault-tolerance \emph{vs.} area trade-off.
Additionally, these designs demand a large number of clock zones leading to higher delay, which is unacceptable for practical realizations.
For example, best existing design of 2-input QCA XOR gate \cite{bhat_INDIACom2015} requires 249 to 529 cells and 6 to 8 clock zones when the above mentioned fault-tolerant schemes are applied on it.
In order to overcome the above shortcomings, we propose new designs of 2-input and 3-input XOR gates in QCA that attain significantly high fault-tolerance with respect to all types of cell misplacement defects. 
The proposed XOR designs requires $85$ and $124$ QCA cells for 2-input and 3-input respectively and hence show significantly better fault-tolerance \emph{vs.} area trade-off than all the existing designs of QCA XOR gates. Furthermore, the absence of any wire crossing (coplanar or multi-layer) in the proposed designs facilitates low cost fabrication. Simulation results are presented based on semiconductor implementation of QCA with an intermediate dot size of about 5$nm$. 

The rest of the paper is organized as follows.
Background and related prior work is discussed in Section \ref{sec:PriorWork}. Section \ref{sec:applicationFTSchemes} demonstrates the applicability of TMR, NAND-multiplexing and majority-multiplexing on existing QCA XOR gates. Design and simulation of the proposed fault-tolerant 2-input and 3-input XOR gates are presented in Section \ref{sec:ProposedDesign}. Detailed comparative study between the proposed XOR gates and the existing ones against various cell misplacement defects is presented in Section \ref{sec:FaultTolerance}. Conclusions are drawn in Section \ref{sec:Conclusions}.

\vspace{-0.15in}
\section{Background and Related Prior Work}
\label{sec:PriorWork}
\begin{figure}[ht]
\center
\begin{minipage}[b]{0.18\linewidth}
\centering
\includegraphics[width=\textwidth]{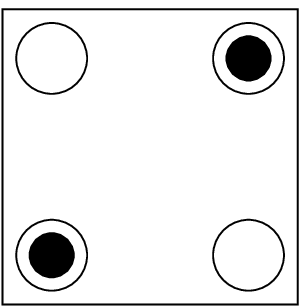}
\centerline{(a)}
\end{minipage}
\hspace{1cm}
\begin{minipage}[b]{0.18\linewidth}
\centering
\includegraphics[width=\textwidth]{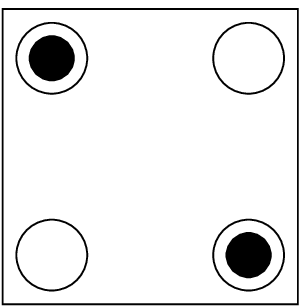}
\centerline{(b)}
\end{minipage}
\caption{QCA cell with (a) $p=+1$ and (b) $p=-1$}
\label{fig:QCACell_Normal}
\end{figure}

The basic element in QCA technology is referred to as {\em cell}. A QCA cell consists of four potential wells or dots located at the four corners of a square. There are two extra electrons which can tunnel quantum mechanically from one dot to another. Due to coulombic repulsion between these two electrons, they always occupy diagonally opposite corner dots. This results in two possible orientations, referred to as polarizations ($p$). Fig. \ref{fig:QCACell_Normal}(a) and Fig. \ref{fig:QCACell_Normal}(b) show QCA cells with $p=+1$ and $p=-1$ respectively. These two polarization states are used to represent binary information 1 and 0 respectively. 

\begin{figure}[ht]
\center
\begin{minipage}[b]{0.18\linewidth}
\centering
\includegraphics[width=\textwidth]{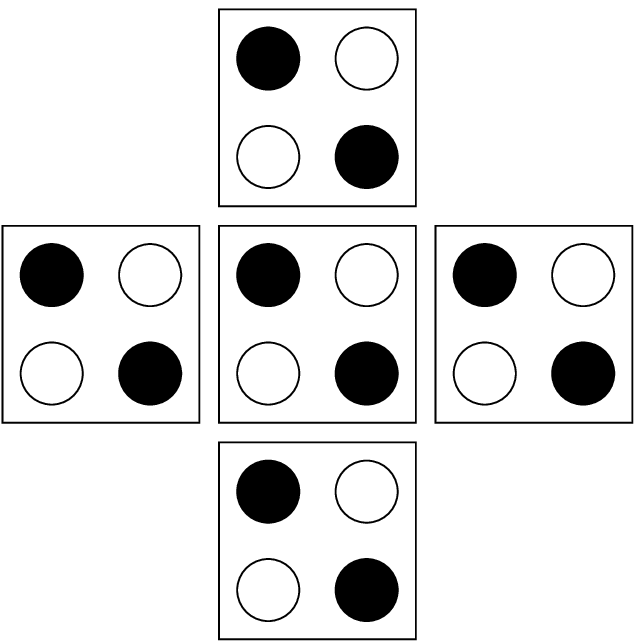}
\centerline{(a)}
\end{minipage}
\hspace{1cm}
\begin{minipage}[b]{0.32\linewidth}
\centering
\includegraphics[width=\textwidth]{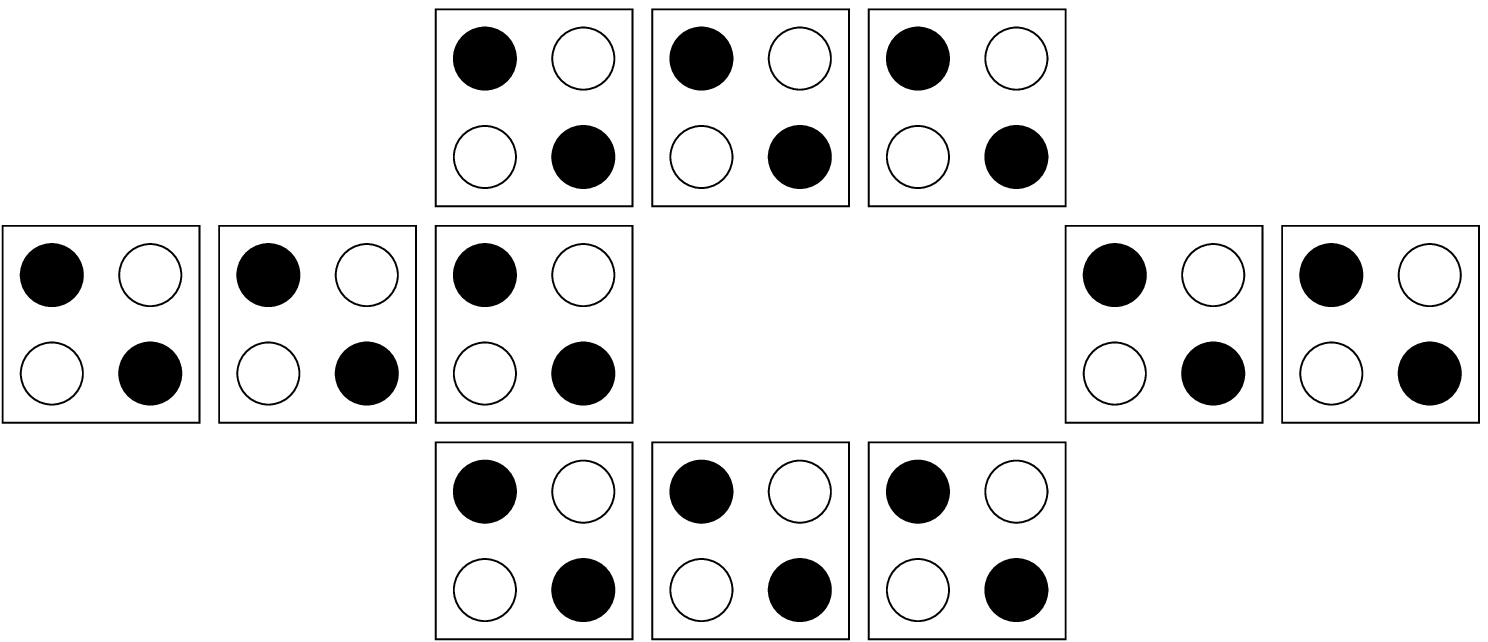}
\centerline{(b)}
\end{minipage}
\caption{Fundamental Gates in QCA (a) Majority Gate (b) Inverter}
\label{fig:Basic_gates}
\end{figure}

\begin{figure}[ht]
\center
\begin{minipage}[b]{0.3\linewidth}
\centering
\includegraphics[width=\textwidth]{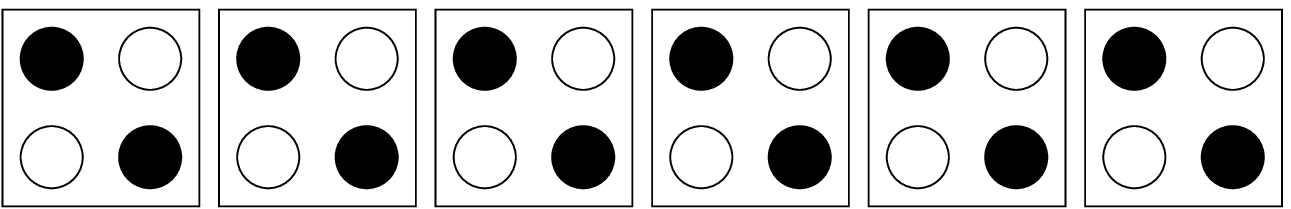}
\centerline{(a)}
\end{minipage}
\hspace{1cm}
\begin{minipage}[b]{0.3\linewidth}
\centering
\includegraphics[width=\textwidth]{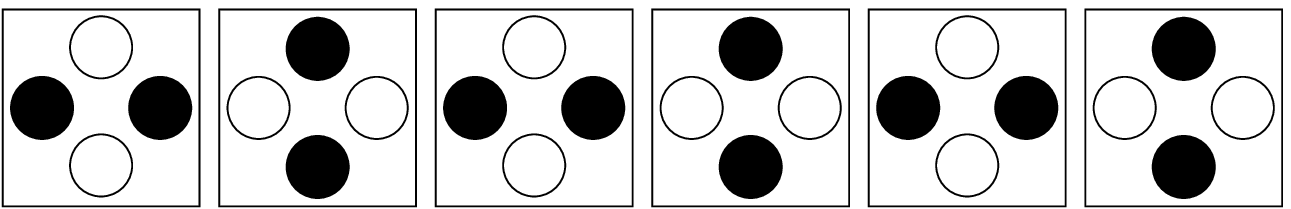}
\centerline{(b)}
\end{minipage}
\caption{Wires in QCA (a) Binary Wire (b) Inverter Chain}
\label{fig:QCA_wires}
\end{figure}

\emph{Majority voter} (MV) and \emph{inverter} gates are considered as the two most fundamental building blocks of QCA. A variety of MVs and inverter gates have been reported in the literature \cite{lent_jap1994, Mitra_MJ_2016, Lombardi_cs2005}.
Typical designs of MV and inverter are shown in Fig. \ref{fig:Basic_gates}(a) and Fig. \ref{fig:Basic_gates}(b) respectively.
% Fig. \ref{fig:Other_Inverter}(a)-(b) show some other possible implementations of QCA inverter.
Two types of QCA wires namely \emph{binary wire} and \emph{inverter chain} are shown in Fig. \ref{fig:QCA_wires}(a) and Fig. \ref{fig:QCA_wires}(b) respectively.

To apply input to a logic device, input cell(s) are forced to assume a particular state by applying external electric field. The input cell(s) then interact with its neighboring cells and change their polarization states accordingly. The process continues and finally the output cell assumes the desired state. Note that the magnitude of the Coulomb force decreases with respect to distance and time. A four-phase clocking scheme \cite{lent_pieee1997} is used to synchronize and control the information flow.

For the last two decades, a major part of the research on QCA has focused on the design and simulation of various digital modules.
Designing different types of adders \cite{Cho_tc2009, swartzlander_asilomar2012, Mitra_MJ_2016} have received considerable interest due to their importance in a computing system.
A conceptual design of QCA XOR gate was first presented in \cite{lent_jap1994}. Note that a straightforward realization of the design is not possible since it does not consider clocking.
A number of QCA implementations of 2-input XOR gate have been presented so far \cite{bhat_INDIACom2015,beigh_cs2013,hema_waset2011,shah_IOSR-JCE2014,beigh_ijpap2013,pallavi_ijsrd2014,Walus_QCADesigner}. A few 3-input QCA XOR gate designs have also appeared in the recent literature \cite{Kianpour_IJNN_2014,Shin_IJCA_2015}.
The designers have used a number of metrics such as area, latency, number of cells, and the type of crossovers. The summary of a comparative study considering these metrics is presented in Section \ref{sec:ProposedDesign} (Table \ref{tab:existingXORcomp} and Table \ref{tab:existingXOR3comp}).     
It is apparent that all the existing designs of 2-input XOR gates show more or less similar performances in terms of number of cells, area and latency with the design proposed by Ahmad and Bhat \cite{bhat_INDIACom2015} seems to be the best. For 3-input XOR, the design presented in \cite{Kianpour_IJNN_2014} outperforms both the designs presented in \cite{Shin_IJCA_2015}.
Interestingly, none of these XOR designs has considered fault-tolerance.
Considering the importance of XOR module in digital circuit design, we analyse the degree of fault-tolerance of these designs. 
Simulation results (Section \ref{sec:FaultTolerance}) show that all of them perform very poorly against most of the cell misplacement defects.
This fact motivate us to investigate the applicability of various existing fault-tolerant schemes in the context of practical realization of QCA XOR gate.
\section{Applicability of Generic fault-tolerant Schemes on Existing QCA XOR Gates}
\label{sec:applicationFTSchemes}
Fault-tolerance of digital circuits is often achieved by adding redundancy into the original design. Triple modular redundancy (TMR) \cite{LyonsTMR1962} and NAND multiplexing \cite{neumann_1956} are the two most popular classical generic fault-tolerance schemes that are often employed to increase the reliability of a circuit module. Recently, a new adaptation of NAND multiplexing referred to as majority multiplexing \cite{roy_tnano2005} has appeared to be a better alternative for nanotechnologies. In this section, we investigate the applicability of TMR, NAND multiplexing, and majority multiplexing in the context of practical realization of fault-tolerant QCA XOR gate.

\begin{figure}[!htb]
\centering
\includegraphics[width=0.3\textwidth]{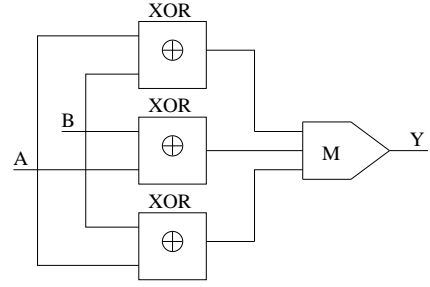}
\caption{\small Block diagram of a TMR system for an XOR gate}
\label{fig:TMR}
\end{figure}

\begin{figure}[!htb]
\centering
%\vspace{-0.1in}
\includegraphics[width=0.44\textwidth]{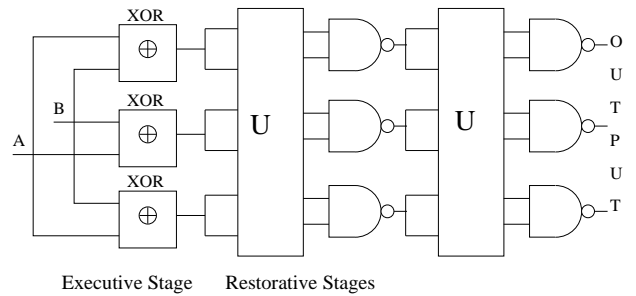}
\caption{\small Block diagram of a NAND multiplexing unit for an XOR gate}
\label{fig:NAND_MUX}
\end{figure}

\begin{figure}[!htb]
\centering
\includegraphics[width=0.45\textwidth]{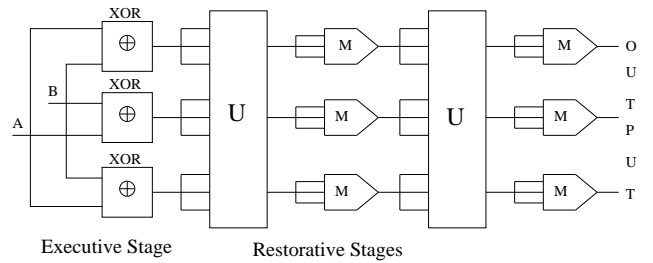}
\caption{\small Block diagram of a majority multiplexing unit for an XOR}
\label{fig:Maj_MUX}
\end{figure}

In a TMR system, the original circuit module is triplicated where the three copies perform the same task in parallel with corresponding outputs being compared through a majority voter circuit. Fig. \ref{fig:TMR} shows the block diagram of a TMR system for a 2-input XOR gate. TMRs can be cascaded to further improve the system's reliability at the cost of higher redundancy.
The basic idea behind NAND multiplexing is to replace the original module by a multiplexing unit, which has N copies of every input and output of the original unit. The multiplexing unit randomly permutates the input signals producing N outputs in parallel. Fig. \ref{fig:NAND_MUX} shows the basic block diagram of a NAND multiplexing system for a 2-input XOR gate. The unit consists of two stages: the {\it executive stage} and the {\it restorative stage}. The executive stage carries out the basic function of the original unit. The restorative stage (consisting of NAND gates) is used to reduce the degradation in the executive stage caused by faults in the original unit. Note that a single stage of NAND gates in the restorative unit inverts the result. Hence, at least two stages are required. The restorative stage can further be iterated to improve the restoration.
Majority multiplexing is similar to NAND multiplexing. As shown in Fig. \ref{fig:Maj_MUX}, the only difference lies with the use of majority gates in place of NAND gates in the restorative stage.
%\vspace{-0.1in}

\vspace{-0.05in}

Although, the generic fault-tolerant techniques discussed above can tolerate a variety of faults (including cell misplacements), the degree of redundancy is unacceptably high.
For example, a single stage TMR system for the best 2-input XOR gate \cite{bhat_INDIACom2015} identified in Section \ref{sec:ProposedDesign} requires minimum 249 QCA cells. 
%Note that in the above calculation, we considered only the cells required to implement %three copies of XOR gate and the majority gate. However, some more cells are required %to complete the desired connections.
Moreover, the fault-tolerance of the TMR system strongly depends on the fault-tolerance capability of the majority gate. Use of a fault-tolerant majority gate \cite{Bibhash_indicon2014} may add up more overhead (22 more cells). The situation become even worse in the case of NAND multiplexing and majority multiplexing.
Construction of a NAND multiplexing system (Fig. \ref{fig:NAND_MUX}) for the same XOR gate \cite{bhat_INDIACom2015} requires minimum 529 cells. Similarly, a majority multiplexing system (Fig. \ref{fig:Maj_MUX}) requires minimum 523 cells. High degree of redundancy involved in all the three systems makes them impractical to realize from both the point of views of area and latency.
The above observations (summarized in Table \ref{tab:existingFTTcomp}) motivate us to design two new practically realizable (in terms of area and delay) fault-tolerant XOR gates which are presented in the next section.
%\vspace{-0.04in}
\begin{table}[htb]
  \begin{center}
    \caption{\small Area and latency incurred by generic fault-tolerant techniques applied on a 2-input QCA XOR gate \cite{bhat_INDIACom2015}}
    \label{tab:existingFTTcomp}
    \begin{tabular}{|l|l|l|l|l|}
      \hline
     
   Technique & Min. & Min. & Latency & Crossover \\
  & \#cells & Area  & (clock  & \\
  &         &      ($\mu m^2$) & phases) &\\
      \hline
    
  TMR & 249 & 0.30 & 6 & Yes  \\
%     & & & & or coplanar\\
      \hline
     
      NAND-MUX & 529 & 0.53 & 8 & Yes\\
      \hline
     
      MAJ-MUX & 523 & 0.40 & 8 & Yes \\
      \hline
    \end{tabular}
  \end{center}
\end{table}
 
\vspace{-0.35in}
\section{Proposed Fault-tolerant XOR Gate Designs}
\label{sec:ProposedDesign}
\begin{figure*}[!htb]
\centering
\includegraphics[width=0.8\textwidth]{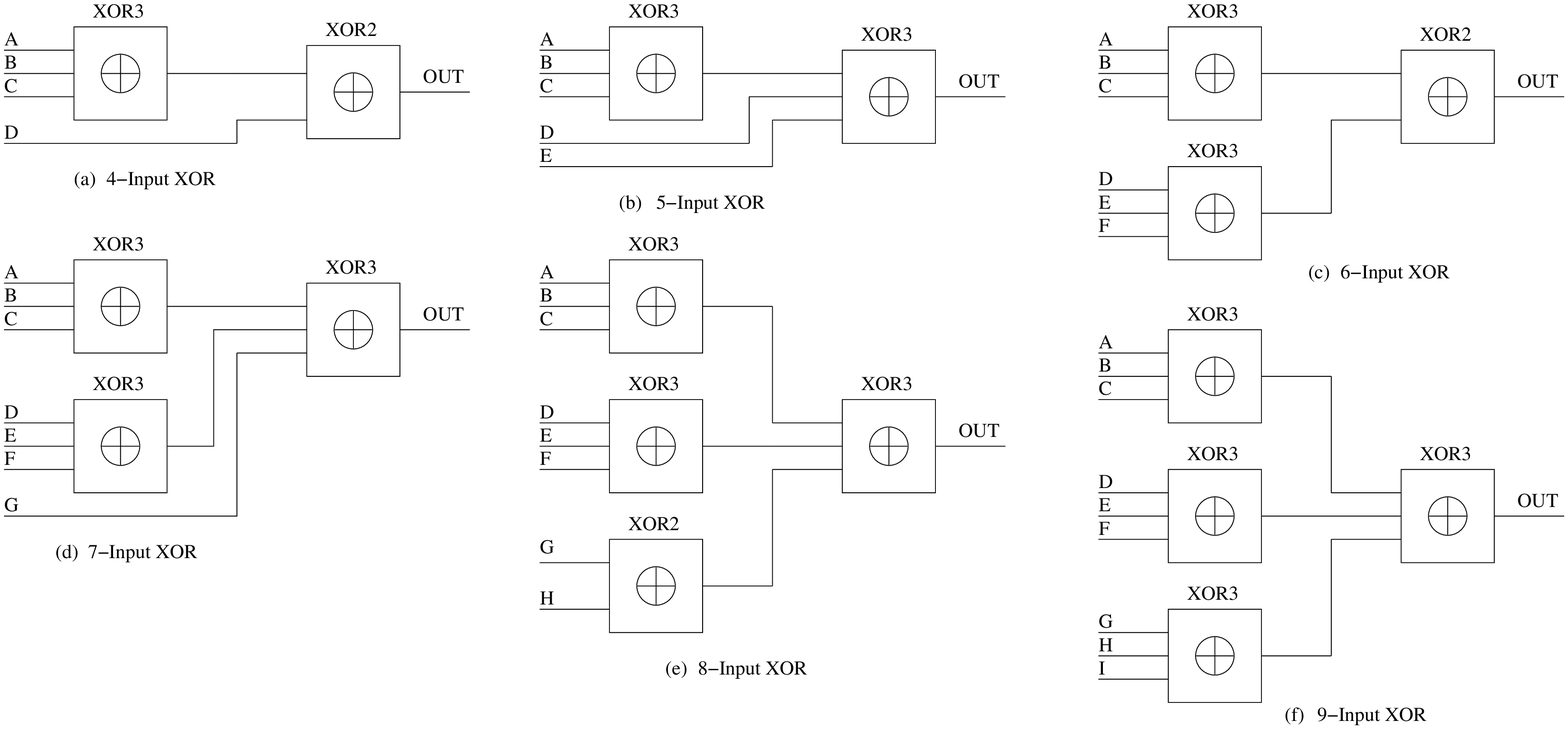}
\caption{\small Gate level implementation of XOR gates of four and higher inputs using only 2 and 3-input XOR gates}
\label{fig:various_xors}
\end{figure*}

As mentioned in Section \ref{sec:Intro}, exclusive-OR operation with two or more variables act as key elements in designing many digital computing subsystems such as adders, linear feedback shift registers, parity generators and checkers etc.
In general, an $n$-variable exclusive-OR function is an odd function defined as the logical sum of the $\frac{2^n}{2}$ minterms whose binary numerical values have an odd number of 1’s \cite{Mano_DigitalBook}.
In conventional CMOS based designs, an XOR gate with three or more inputs is usually constructed by appropriately connecting two or more 2-input XOR gates only. However, a closer look reveals that this approach, especially for QCA based design,  may lead to larger area and delay.
Feinstein and Thornton \cite{Feinstein_RMW2007} have shown how XOR gates with four or higher inputs can be efficiently constructed with the help of 2-input and 3-input XOR gates only. Fig. \ref{fig:various_xors} shows few examples demonstrating the idea.
This motivates us to propose fault-tolerant designs of 2-input and 3-input XOR gates in QCA.

\begin{figure}[!htb]
\centering
\includegraphics[width=0.35\textwidth]{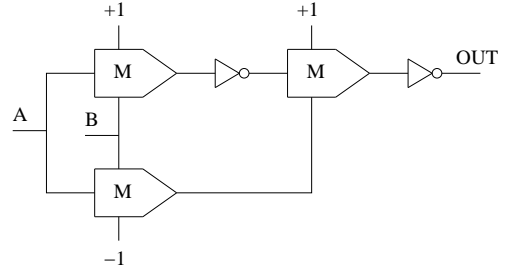}
\caption{\small Gate level implementation of the proposed 2-input XOR gate}
\label{fig:XOR2_block_diagram}
\end{figure}

%\subsection{Fault-tolerant Design of 2-input XOR Gate}
%\label{sec:2inpXOR}
The logical expression representing 2-input XOR function ($A\overline{B}$ + $\overline{A}B$)  can be rewritten equivalently as $(\overline{(\overline{A+B})+AB})$  using simple Boolean algebra.
%A possible gate level implementation of the above expressions is shown in Fig. \ref{fig:XOR_block_diagram}.
As mentioned in Section \ref{sec:PriorWork}, majority gate and inverter pair are commonly used as the basic building blocks in QCA circuits.
Note that a majority gate acts as an AND gate when one of its input cell polarization is set to -1 (logic 0). Similarly, it acts as an OR gate when one of its input cell polarization is set to +1 (logic 1). A NOR gate may be implemented by adding an inverter gate in front of an OR gate. Hence, considering Majority voter and inverter based synthesis (which suits QCA based implementation), the above expression may also be rewritten as $\overline{M[\overline{M(A,B,1)}, M(A, B, 0), 1]}$, where $M(a,b,c)$ represents the majority function defined as $M(a, b, c) = ab + bc + ac$. Figure \ref{fig:XOR2_block_diagram} shows the gate level implementation of the expression.  

%\begin{figure}[!htb]
%\center
%\begin{minipage}[b]{0.47\linewidth}
%\centering
%\includegraphics[width=\textwidth]{Figures/xor2_dia.eps}
%\centerline{(a)}
%\end{minipage}
%\hspace{1cm}
%\begin{minipage}[b]{0.4\linewidth}
%\centering
%\includegraphics[width=\textwidth]{Figures/xor3_dia.eps}
%\centerline{(b)}
%\end{minipage}
%\caption{Gate level implementation of the proposed (a) 2-input XOR gate and (b) 3-input XOR gate}
%\label{fig:XOR_block_diagram}
%\end{figure}

\begin{figure}[!htb]
\centering
\includegraphics[width=0.33\textwidth]{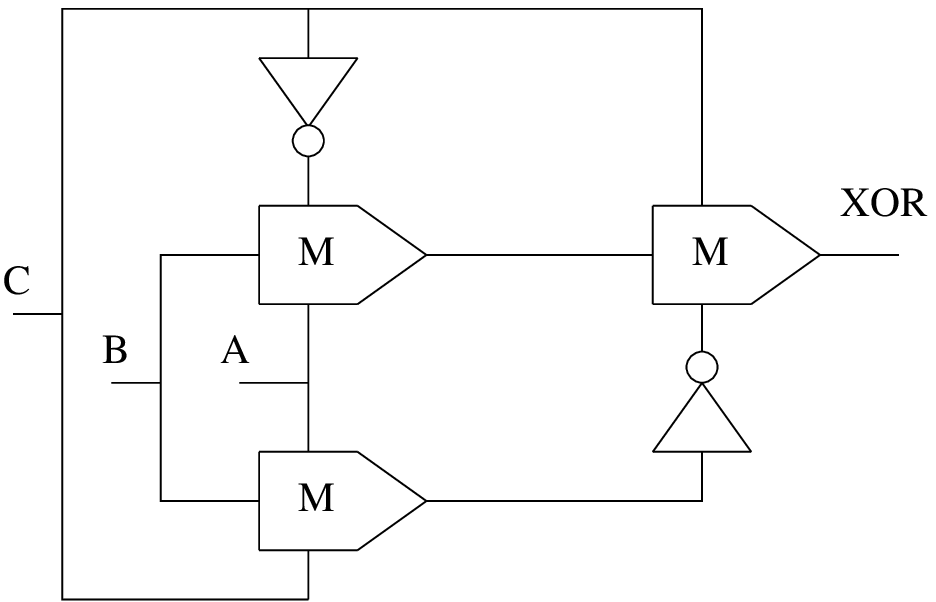}
\caption{\small Gate level implementation of the proposed 3-input XOR gate}
\label{fig:XOR3_block_diagram}
\vspace{-0.1in}
\end{figure}

Following a similar approach, the logical expression representing 3-input XOR function ($\bar{A}\bar{B}C+\bar{A}B\bar{C}+A\bar{B}\bar{C}+ABC$) may be rewritten as $M[\bar{M(A,B,C)}, C, M(A, B,\bar{C})]$.
Figure \ref{fig:XOR3_block_diagram} shows the corresponding gate level implementation.

%%\vspace{-1.6cm}
%\begin{figure}[ht]
%\center
%\begin{minipage}[b]{0.48\linewidth}
%\centering
%\includegraphics[width=\textwidth]{Figures/xor.eps}
%\centerline{(a)}
%\end{minipage}
%%\hspace{-1cm}
%\begin{minipage}[b]{0.43\linewidth}
%\centering
%\includegraphics[width=\textwidth]{Figures/xor3.eps}
%\centerline{(b)}
%\end{minipage}
%\caption{Layout of the proposed  (a) 2-input XOR gate and (b) 3-input XOR gate}
%\label{fig:Proposed_XOR_Layouts}
%\end{figure}

\begin{figure}[htb]
\vspace{-1.0in}
\centering
\includegraphics[width=0.4\textwidth]{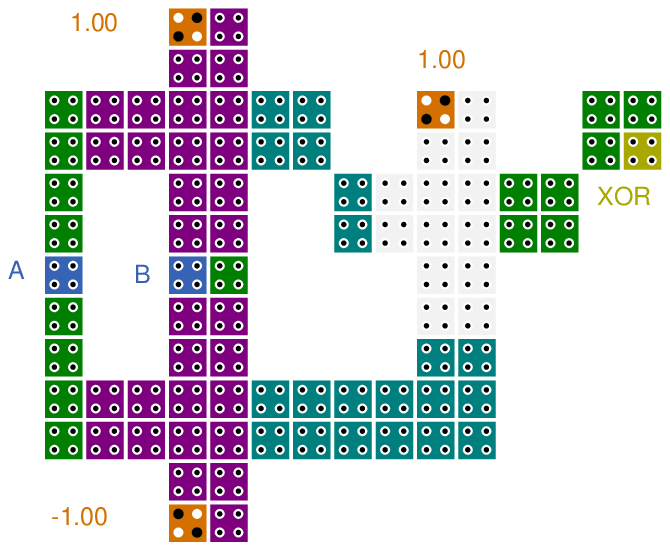}
\vspace{-0.5in}
\caption{Layout of the proposed 2-input XOR gate}
\label{fig:Proposed_XOR2_Layout}
\end{figure}

\begin{figure}[!htb]
\centering
\vspace{-0.6in}
\includegraphics[width=0.4\textwidth]{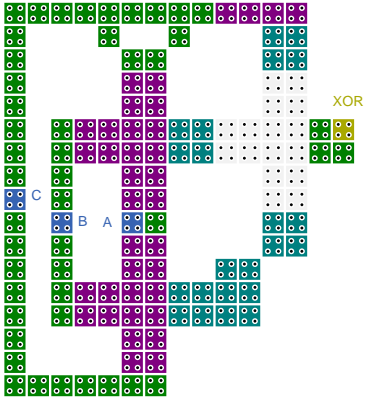}
\vspace{-.3in}
\caption{Layout of the proposed 3-input XOR gate}
\label{fig:Proposed_XOR3_Layout}
\end{figure}
Figure \ref{fig:Proposed_XOR2_Layout} and Figure \ref {fig:Proposed_XOR3_Layout} show the layouts of the proposed 2-input and 3-input XOR gates.
As apparent from the figures, the proposed designs consist of 85 QCA cells and 124 cells respectively.
Assuming QCA cell size of $18nm \times 18nm$ with a gap of $2nm$ between two consecutive cells, the layouts consume area $0.08 \mu m^2$ and $0.10 \mu m^2$ respectively.
Moreover, both the designs use 5 phases (1.25 clock cycles) of clock latency and have no crossover.
Note that a number of redundant cells have been used in the basic building blocks (majority gates, inverters, and connecting wires) of both the designs to improve the fault-tolerance potential of the designs against various types of cell omission and cell misplacement defects.

\begin{figure}[!htb]
\centering
%\vspace{-0.05in}
\includegraphics[width=0.3\textwidth]{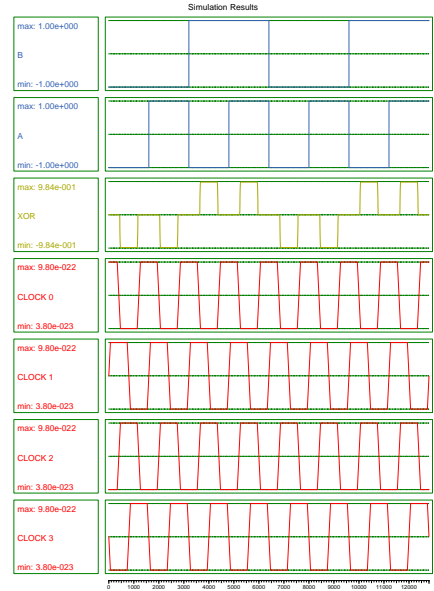}
%\vspace{-0.2in}
\caption{Simulation result for the proposed 2-input XOR gate}
\label{fig:simulation_results}
\end{figure}

\begin{figure}[!htb]
%\vspace{-0.35in}
\centering
\includegraphics[width=0.3\textwidth]{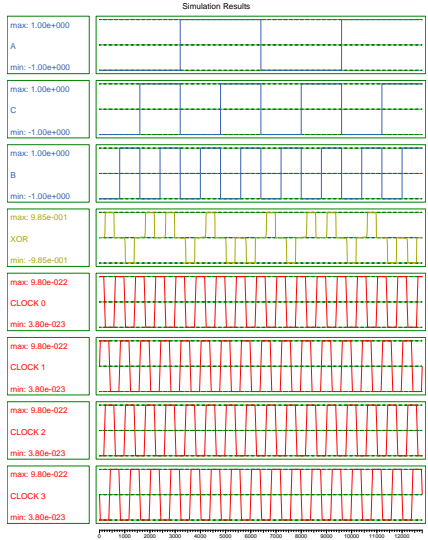}
\caption{Simulation result for the proposed 3-input XOR gate}
\label{fig:simulation_results3}
\end{figure}

To verify the functional behavior of the proposed XOR gates, we carried out simulations with bistable simulation engine of QCADesigner \cite{Walus_QCADesigner} (version 2.0.3) with the following parameters:  (i) Cell size: $18nm \times 18nm$ with a gap of $2nm$ between two consecutive cells, (ii) Radius of effect: $65nm$, (iii) Relative permittivity: 12.9, (iv) Convergence tolerance: 0.001000, (v) Number of samples: 50000, (vi) Intermediate dot size: $5nm$.
The bistable simulation engine of QCADesigner uses intercellular Hartree approximation (ICHA) assuming a simple two-state system to represent each QCA cell.
A little compromise in accuracy as compared to full-basis computation is often compensated by the significantly better scalability \cite{LaRue_TNANO2013}. ICHA is found to be valid and sufficient for verifying the functionality of large QCA circuits.
The simulation output of QCADesigner is shown in Figure \ref{fig:simulation_results} and Figure \ref{fig:simulation_results3} for 2-input and 3-input XOR gates respectively. Note that the maximum polarization at the 2-input XOR output ($P_{max}=0.984$) and at the 3-input XOR output ($P_{max}=0.985$) are significantly strong.
To verify the robustness of the proposed XOR gates further, we have also simulated it for various values of radius of effect (for example, $40nm$ or $75nm$). The behavior of the proposed XORs is found to remain unaltered. The maximum polarization at the output remains almost same.
% The proposed 2-input and 3-input XOR gates occupy an area of $0.078\mu m^2$ and $0.10\mu m^2$ respectively.   

%\vspace{-0.05in}

In order to judge the quality of the proposed designs in terms of the popular design metrics such as area, latency, number of cells, and the type of crossover, we have made a comparative study with existing designs. The summary of the comparative study separately for 2-input and 3-input QCA XOR designs are shown in Table \ref{tab:existingXORcomp} and Table \ref{tab:existingXOR3comp} respectively. It is interesting to note that the proposed 3-input XOR gate outperforms all the existing 3-input XOR designs in terms of the above metrics. Although, the proposed 2-input XOR design cannot outperform the existing designs in terms of these metrics but it does not compromise much too for most of them.
The number of cells used (and hence the area) by the proposed 2-input XOR gate is higher than the most of the existing 2-input XOR designs. However, as described in the next section, the degree of fault-tolerance achieved by the proposed design leads to a significantly better fault-tolerance vs area trade-off as compared to other designs. In fact, fault-tolerance of the proposed designs against cell misplacements is achieved at the cost of redundancy (increased number of cells).  
\begin{table}[h!]
  \begin{center}
%\vspace{-0.06in}
    \caption{Comparisons of various 2-input QCA XOR gates in terms of common design metrics}
    \label{tab:existingXORcomp}
    \begin{tabular}{|l|l|c|r|r|}
      \hline
     
      XOR  & No. of & Area & Latency & Crossover    \\
      
       & \#cells & ($\mu m^2$) & (clock  & \\
       Structures & & & phases) & \\
      \hline
           \cite{hema_waset2011} & 58 & 0.062 & 3 & None \\
      \hline

      \cite{beigh_cs2013} & 42 & 0.036 & 3 & None \\
      \hline
               
      \cite{beigh_ijpap2013} & 41 & 0.044 & 4 & None \\
      \hline

      \cite{pallavi_ijsrd2014} & 62 & 0.090 & 6 & None \\
      \hline
     
     \cite{shah_IOSR-JCE2014} & 35 & 0.040 & 3 & None \\
      \hline

      \cite{bhat_INDIACom2015} & 32 & 0.030 & 3 &  Multilayer \\
      \hline
      
       \cite{Walus_QCADesigner} & 85 & 0.078 & 4 & Multilayer\\
      \hline
      
      Proposed & 85 & 0.078 & 5 & None \\
      
     \hline  
    \end{tabular}
  \end{center}
\end{table}
%\vspace{-0.2in}

\begin{table}[h!]
  \begin{center}
%\vspace{-0.06in}
    \caption{Comparisons of various 3-input QCA XOR gates in terms of common design metrics}
    \label{tab:existingXOR3comp}
    \begin{tabular}{|l|l|c|r|r|}
      \hline
     
      XOR  & No. of & Area & Latency & Crossover      \\
      
       & \#cells & ($\mu m^2$) & (clock  & \\
       Structures & & & phases) & \\
      \hline
       \cite{Kianpour_IJNN_2014} & 98 & 0.12 & 8 & Coplanar \\
      \hline
          \cite{Shin_IJCA_2015} & 164 & 0.22 & 10 & Coplanar  \\
      \hline

      \cite{Shin_IJCA_2015} & 136 & 0.18 & 9 & Coplanar \\
      \hline

     Proposed & 124 & 0.10 & 5 & None \\
     \hline   
    \end{tabular}
  \end{center}
\end{table}
\vspace{-0.2in}
\section{Comparison of proposed XOR gates with existing XORs in terms of fault-tolerance}
\label{sec:FaultTolerance}
In this section, we demonstrate the degree of fault-tolerance achieved by our proposed XOR gates against different cell misplacement defects and also present a detailed comparative study with the existing XOR designs in this regard.
Although none of the existing XOR designs consider fault-tolerance, we reproduce them \cite{bhat_INDIACom2015, beigh_cs2013, hema_waset2011, shah_IOSR-JCE2014, beigh_ijpap2013, pallavi_ijsrd2014, Walus_QCADesigner, Kianpour_IJNN_2014, Shin_IJCA_2015}, and simulate them with bistable simulation engine of QCADesigner \cite{Walus_QCADesigner} (version 2.0.3) to perform the above mentioned comparative study.  
Note that Shin {\it et al.} have proposed two 3-input XOR gates in \cite{Shin_IJCA_2015}.
In this comparative study, we have included the best one (in terms of common design metrics identified in the previous section).

\begin{figure}[htb]
\centering
%\vspace{-0.08in}
\includegraphics[width=0.48\textwidth]{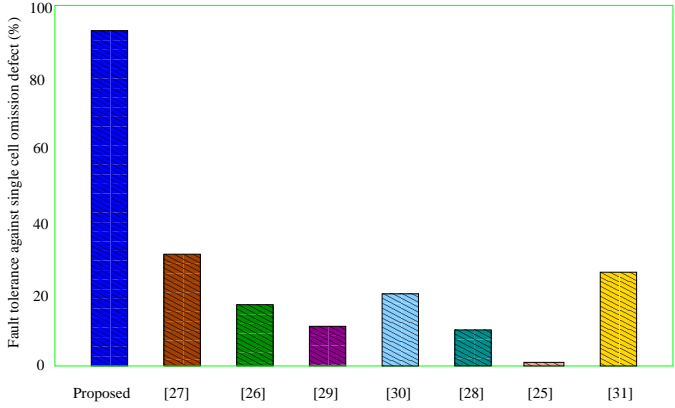}
\caption{Comparison of fault-tolerance of various 2-input XOR gates against single-cell omission defect}
\label{fig:SCO_comparison}
\end{figure}

\begin{figure}[htb]
\centering
%\vspace{-0.08in}
\includegraphics[height=0.35\textwidth]{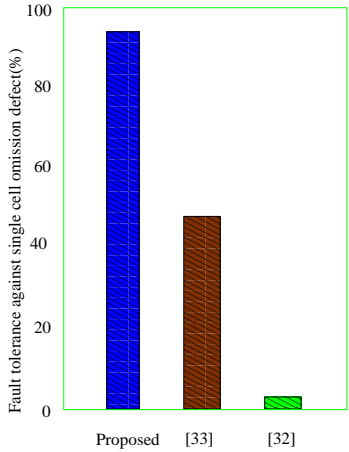}
\caption{Comparison of fault-tolerance of various 3-input XOR gates against single-cell omission defect}
\label{fig:SCO_comparison3}
\end{figure}

%\vspace{-0.08in}
First, we simulate all the XOR gates including the proposed ones for single-cell omission defects at all the device cells (excluding the input/output cells and the cells with fixed polarization). For the proposed 2-input XOR gates, out of 79 device cells omission of 73 of them produce correct output, thereby achieving $92.41\%$ ($\frac{73}{79}\times 100$\%) fault-tolerance.
For the proposed 3-input XOR gates, out of 120 possible instances of single-cell omissions, correct output is produced for 112 cases, thereby achieving $93.33\%$ fault-tolerance.
The summary of comparison is illustrated in Figure \ref{fig:SCO_comparison} and Figure \ref{fig:SCO_comparison3} with the help of bar charts. It is apparent that both the proposed XOR designs outperform their existing counterparts. 

\begin{figure}[!htb]
\centering
\includegraphics[width=0.48\textwidth]{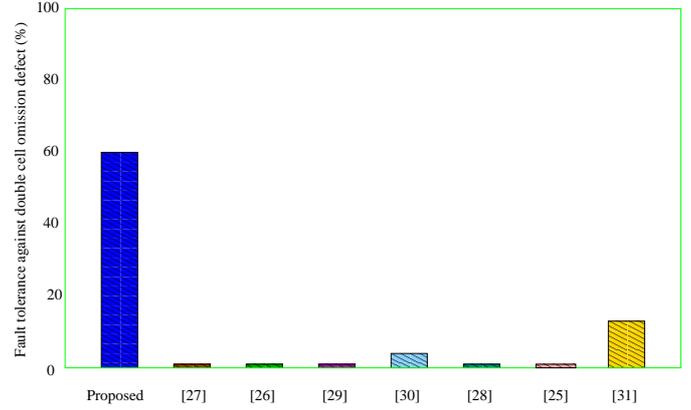}
\caption{Comparison of fault-tolerance of various 2-input XOR gates against double-cell omission defect}
\label{fig:DCO_comparison}
\vspace{0.5cm}
\end{figure}

\begin{figure}[!htb]
\centering
\includegraphics[height=0.35\textwidth]{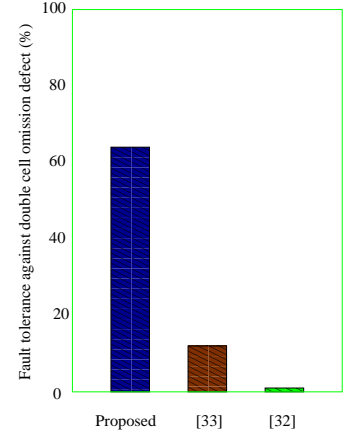}
\caption{Comparison of fault-tolerance of various 3-input XOR gates against double-cell omission defect}
\label{fig:DCO_comparison3}
\end{figure}

We also inspect the effect of double-cell omissions on the functional behavior of various XOR designs. It is found that out of total 107 possible instances of double-cell omissions, the proposed 2-input XOR gate produces correct outputs for 63 cases.
For the proposed 3-input XOR gate, correct output is produced for 100 cases out of total 158 possible instances of double-cell omissions.
Above finding indicate that the proposed XOR gates is able to tolerate double-cell omissions to a large extent.
Simulation results of similar experiments on the existing XOR designs reveal that they can hardly tolerate any double cell omission. This is also apparent from Figures \ref{fig:DCO_comparison} and \ref{fig:DCO_comparison3}.

\begin{figure}[!htb]
\centering
\includegraphics[width=0.48\textwidth]{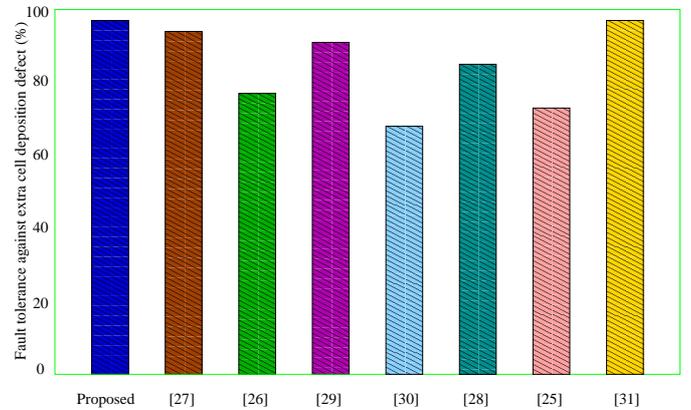}
\caption{Comparison of fault-tolerance of various 2-input XOR gates against extra-cell deposition defect}
\label{fig:ECD_comparision}
\vspace{-0.2in}
\end{figure}

\begin{figure}[!htb]
\centering
\includegraphics[height=0.35\textwidth]{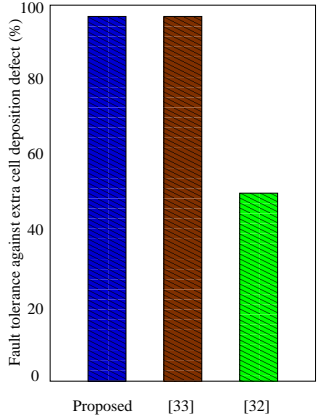}
\caption{Comparison of fault-tolerance of various 3-input XOR gates against extra-cell deposition defect}
\label{fig:ECD_comparision3}
\end{figure}

We next investigate the effect of additional cell deposition defects on the functional behavior of the proposed XOR gates. The output of the proposed 2-input XOR gate is found to produce correct output for 107 cases out of total 111 possible instances of extra cell depositions, thereby achieving $96.40\%$ fault-tolerance. Similarly, the output of the proposed 3-input XOR gate is found to produce correct output for 125 cases out of total 131 possible instances of extra-cell depositions, thereby achieving $95.42\%$ fault-tolerance. The comparison of fault-tolerance achieved by various XOR gates (existing and proposed) against extra-cell deposition defects is illustrated in Figure \ref{fig:ECD_comparision} and Figure \ref{fig:ECD_comparision3}.

\begin{figure*}[!htb]
\centering
\includegraphics[width=\textwidth]{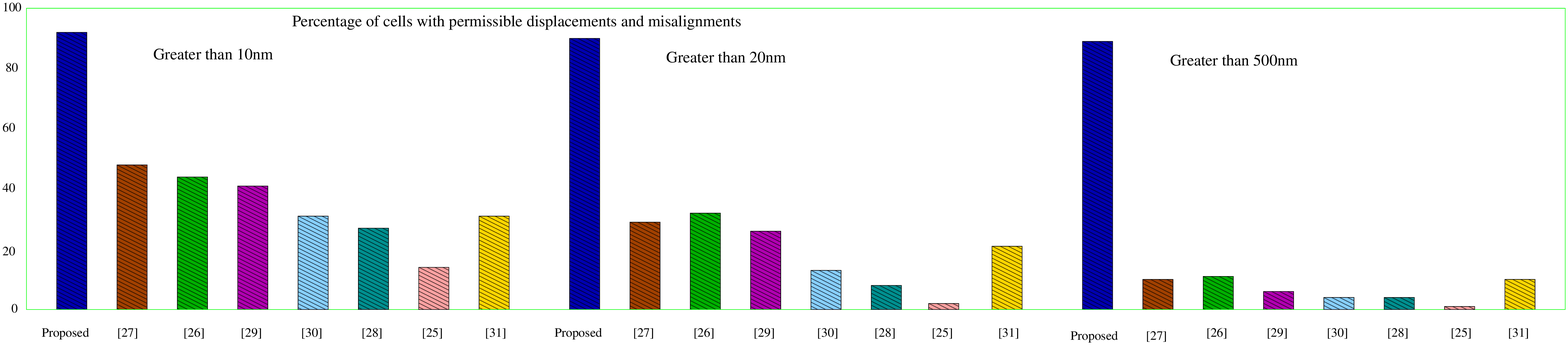}
\caption{Comparison of permissible displacements for various 2-input XOR gates}
\label{fig:CD_comparison}
\end{figure*}

\begin{figure}[!htb]
\centering
\includegraphics[width=0.45\textwidth]{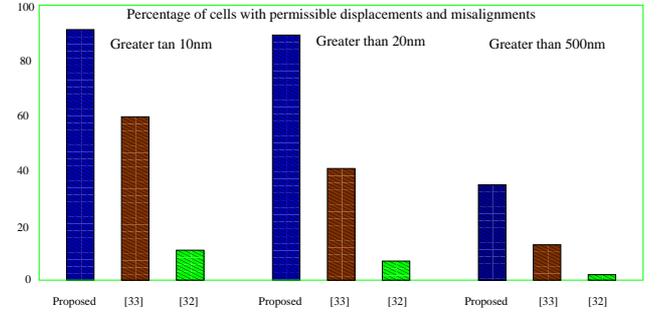}
\caption{Comparison of permissible displacements for various 3-input XOR gates}
\label{fig:CD_comparison3}
\end{figure}

\vspace{1cm}
Finally, we explore the effect of cell displacement and cell misalignment defects on the functional behavior of the proposed XOR gates and all other existing XOR gates.
% The functional behavior of a circuit changes with the magnitude of the displacement and/or misalignment of a cell in a particular direction.
Displacement or misalignment larger than a critical value (referred to as permissible displacement) of a cell in a particular direction causes the circuit to malfunction \cite{Mitra_MJ_2016}.
Larger is the value of permissible displacement associated with a cell displacement/misalignment defect, the circuit is expected to have better fault-tolerance against that defect. The percentage of defects having permissible displacements more than a certain value could be a measure of fault-tolerance of the design against such defects.
We simulate all the XOR designs to find out the percentage of such defects having permissible displacements greater than certain values.
Table \ref{tab:xor_cell_displacements} and Table \ref{tab:xor_cell_displacements3} show the percentage of defects having permissible displacements greater than $10nm$ (i.e., more than the half of the width of a QCA cell), $20nm$ (i.e., more than the width of a QCA cell) and $500nm$ for various 2-input and 3-input XOR gates.
For better illustration, we have also included bar charts to present the comparison (Figures \ref{fig:CD_comparison} and \ref{fig:CD_comparison3}).
It is apparent that the proposed XOR gates completely outperform all the existing ones in terms of fault-tolerance against cell displacement and cell misalignment defects as well.
A significant percentage ($88.54\%$ and $34.78\%$ respectively) having permissible displacements more than $500nm$ for the cell displacement/misalignment defects in the proposed 2-input and 3-input XOR gates indicates that the complete removal of the corresponding cell from the design area does not have any effect on the functional behavior of the circuit.
In other words, the proposed design contains a large number of redundant cells.
In fact, the presence of redundant cells play a major role in achieving higher degree of fault-tolerance against cell omission defects.
\begin{table}[h!]
  \begin{center}
   \scriptsize
    
    \caption{Comparison of fault-tolerance of various 2-input XOR gates against cell displacements and misalignments}
    \label{tab:xor_cell_displacements}
    \begin{tabular}{|l|r|r|r|}
      \hline
      
      XOR Gates & \multicolumn{3}{c|}{Percentage of cells with permissible displacement} \\
      \cline{2-4}

        &  $\geq$ 10nm &  $\geq$ 20nm  &  $\geq$ 500nm  \\
        
      \hline    
        \cite{hema_waset2011} & 47.45 & 29.66 & 10.17  \\
      \hline

      \cite{beigh_cs2013} & 43.90 & 32.93 & 10.98  \\
      \hline
               
      \cite{beigh_ijpap2013} & 41.67 & 26.38 & 5.55  \\
      \hline
       \cite{pallavi_ijsrd2014} & 30.58 & 13.20 & 4.13  \\
      \hline
     
     \cite{shah_IOSR-JCE2014} & 27.78 & 8.33 & 4.16 \\
      \hline

      \cite{bhat_INDIACom2015} & 14.52 & 1.61 & 0.0  \\
      \hline
      
       \cite{Walus_QCADesigner} & 30.89 & 20.78 & 9.55 \\
      \hline
      
      Proposed & 91.67 & 89.58 & 88.54 \\
      \hline
      
     \end{tabular}
  \end{center}
\end{table}

\begin{table}[h!]
  \begin{center}
   \scriptsize
    
    \caption{Comparison of fault-tolerance of various XOR gates against cell displacements and misalignments}
    \label{tab:xor_cell_displacements3}
    \begin{tabular}{|l|r|r|r|}
      \hline
      
      XOR Gates & \multicolumn{3}{c|}{Percentage of cells with permissible displacement} \\
      \cline{2-4}

        &  $\geq$ 10nm &  $\geq$ 20nm  &  $\geq$ 500nm  \\
        
      \hline    
       
          \cite{Shin_IJCA_2015} & 59.35 & 41.73 & 12.95   \\
      \hline  
      \cite{Kianpour_IJNN_2014} & 11.0 & 7.0 & 1.5 \\
      \hline
      
      Proposed & 91.93 & 89.44 & 34.78 \\
      \hline
      
     \end{tabular}
  \end{center}
  \vspace{-.1in}
\end{table}

A consolidated summary of our comparative study on fault-tolerance of various 2-input and 3-input XOR gates (including the proposed XOR gates) against various cell misplacement defects is shown in Table \ref{tab:XOR_FT} and Table \ref{tab:XOR_FT3}. 
\begin{table}[h!]
  \begin{center}
  \scriptsize
    \caption{Comparison summary of fault-tolerance of various 2-input XOR gates against various cell misplacement defects}
    \label{tab:XOR_FT}
    \begin{tabular}{|l|l|l|l|l|}
      \hline
      
     XOR  & \multicolumn{4}{c|}{Percentage of fault-tolerance against} \\
      \cline{2-5}

       Gates &  single-cell & double-cell & extra-cell & cell-displacement  \\

       &  omission &  omission & deposition & \&  misalignment\\ 
     
      \hline
     
       \cite{hema_waset2011} & 30.77 & 0.0 & 94.89 & 47.45  \\
      \hline

      \cite{beigh_cs2013} & 17.14 & 0.0 & 77.0  & 43.90 \\
      \hline
               
      \cite{beigh_ijpap2013} & 11.43 & 0.0 & 91.3 & 41.67 \\
      \hline
      
       \cite{pallavi_ijsrd2014} & 20.0 & 3.70 & 67.54 & 30.58 \\
      \hline
     
     \cite{shah_IOSR-JCE2014} & 10.34 & 0.0 & 85.71 & 27.78\\
      \hline

      \cite{bhat_INDIACom2015} & 0.0 & 0.0 & 73.68  & 14.52 \\
      \hline
      
       \cite{Walus_QCADesigner} & 26.58 & 12.86 & 97.20 & 30.89 \\
      \hline
      
       Proposed & 92.31 & 58.90 & 96.40 & 91.67  \\
 %      Paper &&&&\\
      \hline
         
    \end{tabular}
  \end{center}
\end{table}
%\vspace{-0.1in}

\begin{table}[h!]
  \begin{center}
  \scriptsize
    \caption{Comparison summary of fault-tolerance of various 3-input XOR gates against various cell misplacement defects}
    \label{tab:XOR_FT3}
    \begin{tabular}{|l|l|l|l|l|}
      \hline
      
     XOR  & \multicolumn{4}{c|}{Percentage of fault-tolerance against} \\
      \cline{2-5}

       Gates &  single-cell & double-cell & extra-cell & cell-displacement  \\

       &  omission &  omission & deposition & \&  misalignment\\ 
     
      \hline
     
       \cite{Shin_IJCA_2015} & 48.41 & 11.90 & 95.40 & 59.35  \\
      \hline
               
      \cite{Kianpour_IJNN_2014} & 2.27 & 0.0 & 50.0 & 11.0 \\
      \hline
      
       Proposed & 93.33 & 63.29 & 95.42 & 91.93  \\
%        Paper &&&&\\
      \hline
         
    \end{tabular}
  \end{center}
  \vspace{-0.2in}
\end{table}

\vspace{-0.2in}
\section{Conclusions}
\label{sec:Conclusions}
%The conclusion goes here.
Design of effective fault-tolerant schemes are desirable for reliable realization of various digital modules in QCA.
XOR gates are found to be one of the important component used in the construction of several computing subsystems. 
%emerging nanotechnologies such systems. 
The applicability of popular fault-tolerant schemes such as TMR, NAND multiplexing, and majority multiplexing in the context of practical realization of QCA XOR gate has been investigated and is observed to perform poorly. In order to bridge the gap a new fault-tolerant designs of QCA XOR gates have been presented. Simulation results show that the proposed designs achieve significantly high fault-tolerance against various cell misplacement defects and completely outperform existing counterparts in this regard. Absence of any crossover in the physical layout of the proposed gates further enhances practical realizability of the designs. 

\vspace{-0.1in}
\section*{Acknowledgment}
The authors would like to thank Dr. Bibhash Sen for his valuable suggestions.
%\vspace{-0.15in}

%\section*{References}
\vspace{-0.12in}
\section*{References}
\vspace{-0.12in}
\bibliography{FT_XOR}

\end{document}